%% file: main.tex
\documentclass[journal=jctcce,manuscript=article
,layout=twocolumn
]{achemso}
\usepackage[fontsize=10pt]{fontsize}
\usepackage{chemformula} % Formula subscripts using \ch{}
\usepackage[T1]{fontenc} % Use modern font encodings
\usepackage{amsmath}
\usepackage{appendix}
\usepackage{booktabs}

\author{Richard Einsele}
\author{Roland Mitri\'c}
\email{roland.mitric@uni-wuerzburg.de}
\phone{+49 (0)931 31 85135}
\affiliation{Institut für Physikalische und Theoretische Chemie, Julius-Maximilians-Universität,  Würzburg, Germany}

\title
  {Nonadiabatic excited-state dynamics and energy gradients in the framework of FMO-LC-TDDFTB}


\let\oldmaketitle\maketitle
\let\maketitle\relax

\begin{document}

\begin{tocentry}
    \includegraphics{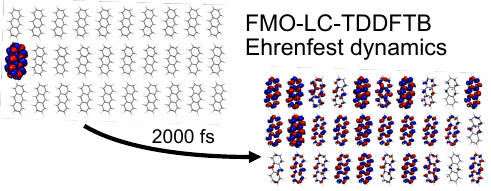}
\end{tocentry}

\twocolumn[
\begin{@twocolumnfalse}
\oldmaketitle
\begin{abstract}
We introduce a novel methodology for simulating the excited-state dynamics of extensive molecular aggregates in the framework of the long-range corrected time-dependent density-functional tight-binding fragment molecular orbital method (FMO-LC-TDDFTB) combined with the mean-field Ehrenfest method. The electronic structure of the system is described in a quasi-diabatic basis composed of locally excited and charge-transfer states of all fragments. In order to carry out nonadiabatic molecular dynamics simulatios, we derive and implement the excited-state gradients of the locally excited and charge-transfer states. Subsequently, the accuracy of the analytical excited-state gradients is evaluated.
The applicability to the simulation of exciton transport in organic semiconductors is illustrated on a large cluster of anthracene molecules. Additionally, nonadiabatic molecular dynamics simulations of a model system of benzothieno-benzothiophene molecules highlight the method's utility in studying charge-transfer dynamics in organic materials. Our new methodology will facilitate the investigation of excitonic transfer in extensive biological systems, nanomaterials and other complex molecular systems consisting of thousands of atoms.  
\end{abstract}
\end{@twocolumnfalse}
]

\input{introduction}
\input{theory}
\input{results}
\input{conclusion}

\begin{acknowledgement}
We gratefully acknowledge financial support by the Deutsche Forschungsgemeinschaft via the grants MI1236/6-1 and MI1236/7-1.
\end{acknowledgement}

% %%%%%%%%%%%%%%%%%%%%%%%%%%%%%%%%%%%%%%%%%%%%%%%%%%%%%%%%%%%%%%%%%%%%%
% %% The same is true for Supporting Information, which should use the
% %% suppinfo environment.
% %%%%%%%%%%%%%%%%%%%%%%%%%%%%%%%%%%%%%%%%%%%%%%%%%%%%%%%%%%%%%%%%%%%%%
% \begin{suppinfo}

% A listing of the contents of each file supplied as Supporting Information
% should be included. For instructions on what should be included in the
% Supporting Information as well as how to prepare this material for
% publications, refer to the journal's Instructions for Authors.

% The following files are available free of charge.
% \begin{itemize}
%   \item Filename: brief description
%   \item Filename: brief description
% \end{itemize}

% \end{suppinfo}

\input{appendix}
\newpage
\bibliography{references}

\end{document}

%% file: introduction.tex
\section{Introduction}
The advancement of quantum chemical methodologies within the paradigm of excited-state molecular dynamics is imperative for probing excited-state phenomena within bio-molecules, organic semiconductors, or nanomaterials.

Whereas ab-initio wave-function or density-functional theory (DFT) methods may be utilized to simulate excited-state molecular dynamics in moderately sized molecular systems, the calculation of larger systems is accompanied by a huge increase in computational time due to the typically observed cubic or higher scaling of these methods.

The development of semiempirical quantum mechanical (SQM) methodologies has enabled the simulation of molecular dynamics in larger systems with reduced computational demands, owing to the utilization of minimal basis sets and the neglect of differential overlap between atomic basis functions \cite{thiel_semiempirical_2014,christensen_semiempirical_2016}. Notably, among the spectrum of SQM techniques, the relatively recent emergence of the density-functional tight-binding (DFTB)\cite{porezag_construction_1995,seifert_calculations_1996,elstner_self-consistent-charge_1998,seifert_densityfunctional_2012} method has demonstrated efficacy in simulating processes within the excited-state manifold\cite{niehaus_tight-binding_2001,gao_nonadiabatic_2014,ilawe_real-time_2017,stojanovic_nonadiabatic_2017,titov_exciton_2018,posenitskiy_non-adiabatic_2019,inamori_spinflip_2020,uratani_fast_2020,wu_nonadiabatic_2022}. Furthermore, the extension of the DFTB method with the long-range correction has enabled the investigation of charge-transfer excitations\cite{humeniuk_long-range_2015}. The long-range corrected time-dependent density-functional tight-binding (LC-TD-DFTB) method has been integrated into multiple software packages, such as DFTB+\cite{hourahine_dftb_2020}, DFTBaby\cite{humeniuk_dftbaby_2017}, GAMESS\cite{barca_recent_2020,zahariev_general_2023} and DIALECT\cite{einsele_long-range_2023}, enabling the investigation of excited-state dynamics in molecular systems comprising several hundred atoms.
Another semiempirical tight-binding approach is the gfn-xtb methodology of Grimme and coworkers\cite{grimme_robust_2017,bannwarth_gfn2-xtbaccurate_2019,bannwarth_extended_2021}, which shows excellent results for ground-state molecular properties and dynamics simulations. Although the method has been employed to screen minimum energy crossing points between different electronic states\cite{pracht_fast_2022,bannwarth_sqmbox_2023}, the gfn-xtb method has not yet been expanded to describe the excited-state dynamics of extensive molecular systems.

Hybrid schemes, exemplified by QM/MM methods\cite{dapprich_new_1999,vreven_combining_2006}, present an additional approach for effectively simulating the excited-state dynamics of larger molecular systems. The integration of quantum mechanical treatment for important segments of the molecular systems with classical treatment for the remaining parts yields a significant reduction in computational demand, while upholding precise characterization of the quantum region. Nevertheless, these methodologies constrain the description of the excited-state manifold to a subset of the entire molecular system, thereby diminishing the approach's versatility.

The Multi-Configuration Time-Dependent Hartree (MCTDH)\cite{meyer_multi-configurational_1990,manthe_wavepacket_1992,beck_multiconfiguration_2000} method and its multilayer derivative (ML-MCTDH)\cite{wang_multilayer_2003,wang_multilayer_2015,vendrell_multilayer_2011} are other promising alternatives for the investigation of the excited-state dynamics of large molecular systems. They have been extensively used to study the exciton transport in organic and nanomaterials.\cite{binder_ultrafast_2017,binder_conformational_2018,binder_first-principles_2019,popp_coherent_2019,binder_first-principles_2020,di_maiolo_quantum_2020,hegger_first-principles_2020,popp_quantum_2021,mondelo-martell_quantum_2022,brey_coherent_2024}

To model the molecular dynamics of several thousand atoms, linear scaling methodologies such as the fragment molecular orbital (FMO)\cite{kitaura_fragment_1999,nakano_fragment_2000,nakano_fragment_2002,fedorov_multilayer_2005} method and the divide-and-conquer (DC)\cite{yang_density-matrix_1995,akama_is_2007,akama_implementation_2007,kobayashi_divide-and-conquer_2010,nakai_divide-and-conquer_2023} method have been devised. These techniques employ fragmentation schemes, wherein the molecular system is divided into multiple segments, and the properties of the entire system are derived from the interactions among the isolated fragments. These approaches have been implemented within various quantum chemical methods, including time-dependent density-functional theory (TDDFT)\cite{chiba_time-dependent_2007,chiba_time-dependent_2007-1,nakata_analytic_2023}, configuration interaction (CI)\cite{mochizuki_configuration_2005,fujita_development_2018}, the coupled-cluster method\cite{kobayashi_extension_2008,yoshikawa_large-scale_2020} and the GW approximation\cite{fujita_fragment-based_2021}. Nevertheless, integrating fragmentation strategies with semiempirical quantum mechanical methods yields a further increase in computational efficiency. Consequently, the combination of the fragment molecular orbital and divide-and-conquer schemes with the density-functional tight-binding method (FMO-DFTB, DC-DFTB) has enabled the simulation of molecular dynamics in various materials such as organic semiconductors or peptides in their electronic ground states\cite{nishimoto_density-functional_2014,nishimoto_large-scale_2015,nishizawa_three_2016,nishimura_parallel_2018}. 
The FMO-DFTB method has also been expanded to incorporate the polarizable continuum model\cite{nishimoto_fragment_2016}, the long-range correction (FMO-LC-DFTB)\cite{vuong_fragment_2019}, periodic boundary conditions\cite{nishimoto_fragment_2021} and polarization effects\cite{fedorov_polarization_2022}.

While these fragmentation approaches have been utilized for exploring excited-state molecular dynamics, their utility has hitherto been restricted to molecular systems where the excitation was confined to a limited subset of molecules, with the remainder of the system treated in its electronic ground state\cite{komoto_development_2019,nishimura_dcdftbmd_2019,komoto_large-scale_2020,uratani_non-adiabatic_2020,uratani_trajectory_2021,nakata_analytic_2023}. In our previous work, we combined the FMO-LC-DFTB method with an excitonic Hamiltonian to calculate the complete excited-state spectrum of large molecular aggregates (FMO-LC-TDDFTB)\cite{einsele_long-range_2023}. To this end, we employed a quasi-diabatic basis that consists of locally excited (LE) and charge-transfer (CT) states, which are obtained from LC-TD-DFTB calculations of monomer and pair fragments. The excitonic Hamiltonian is then built from the energies of the basis states and the electronic couplings between them. Subsequently, the excited-state energies of the molecular system are obtained by the diagonalization of the excitonic Hamiltonian.

In the present work, we further extend our FMO-LC-TDDFTB methodology to include dynamical processes. Given the considerable number of electronically excited states within extensive molecular systems, our new theoretical framework employs the mean-field Ehrenfest\cite{ehrenfest_bemerkung_1927,sawada_mean-trajectory_1985,li_ab_2005} method instead of a Surface Hopping\cite{tully_trajectory_1971,barbatti_nonadiabatic_2011} approach. Through the utilization of the Ehrenfest methodology within the quasi-diabatic basis of LE and CT states, we describe the excited-state dynamics of extensive molecular systems including all nuclear degrees of freedom. We derive the analytical gradients for the LE and CT states and check their accuracy by comparing the analytical values to the numerical results. Subsequently, we investigate the excited-state exciton dynamics of a two-dimensional anthracene structure, followed by the simulation of the charge-transfer dynamics of a model system of a p-type organic semiconductior, [1]benzothieno[3,2‐b]benzothiophene (BTBT)\cite{ebata_highly_2007,xie_structures_2022}.  

This paper is structured as follows: In Section \ref{sec:methodology}, the theoretical framework of our method is defined and the technicalities of the molecular dynamics simulations are described. The accuracy of the quasi-diabatic excited-state gradients and the results of the excited-state Ehrenfest molecular dynamics simulations are shown in Section \ref{sec:results}. In section \ref{sec:conclusion}, conclusions and an outlook are given.

%% file: theory.tex
\newcommand\numberthis{\addtocounter{equation}{1}\tag{\theequation}}
\section{Methodology}\label{sec:methodology}
In this chapter, we will derive the expressions for the gradients of our FMO-LC-TDDFTB method. At first, we give a brief summary of the FMO-DFTB formalism for the electronic ground state \cite{nishimoto_density-functional_2014}. Subsequently, the methodology of the ground-state gradient of the FMO-DFTB method is presented. In addition, we summarize the theory behind the FMO-LC-TDDFTB method. This is followed by the derivation of the gradients for the quasi-diabatic LE and CT states in the framework of the FMO-LC-TDDFTB method. Subsequently, the methodology of the excited-state Ehrenfest molecular dynamics in the basis of the locally excited and charge-transfer states is described. Finally, the computational technicalities of the molecular dynamics simulations are provided.

In this manuscript, atomic units are employed, accompanied by the following notation convention: uppercase letters A and B represent atoms, while molecular fragments are denoted by uppercase letters I through L without indices. Matrix elements are indicated by uppercase letters with indices, and matrices are represented by bold uppercase letters. Molecular orbital indices are denoted by lowercase letters (Greek letters signify atomic orbital indices).
\subsection{FMO-LC-DFTB}
\textbf{Ground state energies}

After Nishimoto \textit{et al}\cite{nishimoto_density-functional_2014} firstly combined the fragment molecular orbital method with the density functional tight-binding formalism, it was extended by Vuong \textit{et al} to include LC-DFTB\cite{vuong_fragment_2019}. We used this theory as the foundation for the ground-state calculation of our FMO-LC-TDDFTB method. In this section, we only give a short summary of the working equations of FMO-LC-DFTB, as it is described in detail in Refs. \cite{nishimoto_density-functional_2014,vuong_fragment_2019,einsele_long-range_2023}.
The total ground-state energy of the complete system in FMO-LC-DFTB is given by
\begin{align}\label{eq:fmo_ground_state}
\begin{split}
    E=&\sum_I^N E_I+ \sum_I^N \sum_{J > I}^N\left(E_{I J}-E_I-E_J\right) \\
    &+ \sum_I^N \sum_{J > I}^N \Delta E_{I J}^{\mathrm{em}},
\end{split}
\end{align}
where
\begin{align}
\begin{split}
    E_X =& \sum_{\mu \nu} P_{\mu v} H_{\mu v}^0 \\
 &+\frac{1}{2} \sum_{\mu, \sigma, \lambda, v} \Delta P_{\mu \sigma} \Delta P_{\lambda \nu}(\mu \sigma \mid \lambda v)  \\
& -\frac{1}{4} \sum_{\mu, \sigma, \lambda, v} \Delta P_{\mu \sigma} \Delta P_{\lambda v}(\mu \lambda \mid \sigma v)_{\mathrm{lr}} \\
 &+\sum_{A, B} V_{A B}^{\mathrm{rep}}\left(R_{A B}\right)
\end{split}
\end{align}
is the internal LC-DFTB energy of a single fragment of the system. While $E_{I}$ ($E_{J}$) denotes the energy of a monomer fragment, $E_{IJ}$ denotes the energy of a pair fragment. Using the tight-binding formalism, the two-electron integrals in the Coulomb and exchange energy contributions  can be expressed as 
\begin{align*}\label{eq:2e_approx_ao}
(\mu \lambda \mid \sigma \nu) &=\iint \phi_{\mu}(r_1) \phi_{\lambda}(r_1) \frac{1}{r_{12}} \phi_{\sigma}(r_2) \phi_{\nu}(r_2) \mathrm{d} 1 \mathrm{d} 2 \\
& \approx \sum_{A, B} \gamma_{A B} q_{A}^{\mu \lambda} q_{B}^{\sigma \nu}, \numberthis
\end{align*}
and 
\begin{align*}\label{eq:2e_approx_ao_lr}
(\mu \lambda \mid \sigma \nu)_{lr} &=\iint \phi_{\mu}(r_1) \phi_{\lambda}(r_1) \frac{\mathrm{erf}(\frac{r_{12}}{R_{lr}})}{r_{12}} \phi_{\sigma}(r_2) \phi_{\nu}(r_2) \mathrm{d} 1 \mathrm{d} 2 \\
& \approx \sum_{A, B} \gamma_{A B}^{lr} q_{A}^{\mu \lambda} q_{B}^{\sigma \nu}, \numberthis
\end{align*}
where the $\gamma$ and $\gamma^{lr}$ matrices, which represent charge fluctuation interactions are defined in Ref.\cite{einsele_long-range_2023} and
the transition charges in the AO basis $q_{A}^{\mu \lambda}$ of atom A are
\begin{equation}
q_{A}^{\mu \lambda}=\frac{1}{2}(\delta(\mu \in A)+\delta(\lambda \in A)) S_{\mu \nu}.
\label{eq:q_between_aos}
\end{equation}
The last term of Eq. (\ref{eq:fmo_ground_state}) is the difference of the embedding energy of the pair and the monomers
\begin{align}\label{eq:embedding_energy}
\begin{split}
    \Delta E_{I J}^{\mathrm{em}}&=E_{I J}^{\mathrm{em}}-E_I^{\mathrm{em}}-E_J^{\mathrm{em}}\\
    &=\sum_{A \in I J} \sum_{K \neq I, J}^N \sum_{C \in K} \gamma_{A C} \Delta \Delta q_A^{I J} \Delta q_C^K,
\end{split}
\end{align}
which describes the difference in the Coulomb interaction between the molecular environment and the respective fragments. While the term
\begin{align}
\Delta q_A & =q_A-q_A^0 \\
& =\sum_{\mu \in A} \sum_v\left[P_{\mu v} S_{\mu v}-P_{\mu v}^0 S_{\mu v}\right]
\end{align}
is the Mulliken charge difference of atom A, $\Delta \Delta q_A^{I J}$ is the difference in the Mulliken charges of the pair and the respective monomers. 

We use the electrostatic-dimer (ES-DIM)\cite{nakano_fragment_2002} approximation to account for far-separated pairs that have zero orbital overlap. The energy of ES-DIM pairs is given by
\begin{equation}\label{eq:ESDIM_energy}
    E_{I J}=E_I+E_J+\sum_{A \in I} \sum_{B \in J} \gamma_{A B} \Delta q_A^I \Delta q_B^J.
\end{equation}
As the FMO-DFTB method includes the Coulomb interaction between all monomer fragments 
\begin{equation}
    V_{\mu v}^X=\frac{1}{2} S_{\mu \nu}^X \sum_{K \neq X}^N \sum_{C \in K}^{N_K}\left(\gamma_{A C}+\gamma_{B C}\right) \Delta q_C^K
\end{equation}
in the ground-state Hamiltonian of a fragment,
\begin{align}
\begin{split}
    H_{\mu \nu}^X&=H_{\mu \nu}^{0}+\frac{1}{2} S_{\mu \nu} \sum_{C}\left(\gamma_{A C}+\gamma_{B C}\right) \Delta q_{C} \\
&+V_{\mu \nu}^X -\frac{1}{8} \sum_{\alpha \beta} \Delta P_{\alpha \beta} S_{\mu \alpha} S_{\beta \nu} \\
& \times \left(\gamma_{\mu \beta}^{\mathrm{lr}}+\gamma_{\mu \nu}^{\mathrm{lr}}+\gamma_{\alpha \beta}^{\mathrm{lr}}+\gamma_{\alpha \nu}^{\mathrm{lr}}\right),
\end{split}
\end{align}
the FMO-DFTB ground states calculation requires simultaneous monomer self-consistent charge (SCC) iterations, where $V_{\mu v}^X$ is updated in each iteration. A detailed description of the technicalities of the implementation of the monomer SCC iterations is given in our previous work.\cite{einsele_long-range_2023}

\textbf{Ground-state gradients}
\newline
In the following, a short summary of the derivation of the ground-state gradient of the FMO-LC-DFTB method is presented, which was originally published in Refs. \cite{nishimoto_density-functional_2014,vuong_fragment_2019}. The total gradient of the system 
\begin{align}\label{eq:fmo_gs_grad}
\begin{split}
    \frac{\partial E}{\partial r_a}&=\sum_I^N \frac{\partial E_I}{\partial r_{\alpha x}}+\sum_{I>J}^N\left(\frac{\partial E_{I J}}{\partial r_{\alpha x}}-\frac{\partial E_I}{\partial r_{\alpha x}}-\frac{\partial E_J}{\partial r_{\alpha x}}\right)\\
    &+\sum_{I>J}^N \frac{\partial \Delta E_{I J}^{em}}{\partial r_{\alpha x}}
\end{split}
\end{align}
is composed of the derivatives of the various energy expressions (\textit{cf}. Eq. (\ref{eq:fmo_ground_state})).
The energy gradient of a monomer or pair fragment
\begin{align*}
\frac{\partial E_X}{\partial r_a} & =\sum_{(b \neq a) \in X} \sum_{\mu \in a} \sum_{\nu \in b}\left[2 P_{\mu \nu}^X \frac{\partial H_{\mu \nu}^{0, X}}{\partial r_a}-2 W_{\mu \nu}^{\prime X} \frac{\partial S_{\mu \nu}^X}{\partial r_a} \right.\\
& +P_{\mu \nu}^X \frac{\partial S_{\mu \nu}^X}{\partial r_a} \sum_{c \in X}\left(\gamma_{a c}+\gamma_{b c}\right) \Delta q_c^X-\frac{1}{4} \frac{\partial S_{\mu \nu}^X}{\partial r_a} \\
& \times \sum_{\alpha \beta \in X} \Delta P_{\mu \alpha}^X \Delta P_{\beta \nu}^X S_{\alpha \beta}^X\left(\gamma_{a \alpha}^{\mathrm{lr}}+\gamma_{a \beta}^{\mathrm{lr}}+\gamma_{b \alpha}^{\mathrm{lr}}+\gamma_{b \beta}^{\mathrm{lr}}\right)  \\
& -\frac{1}{4} \frac{\partial \gamma_{a b}^{\mathrm{lr}}}{\partial r_a} \sum_{\alpha \beta \in X} S_{\mu \beta}^X S_{\nu \alpha}^X  \numberthis \\ 
& \times \left(\Delta P_{\mu \alpha}^X \Delta P_{\beta \nu}^X
 +\Delta P_{\mu \nu}^X \Delta P_{\alpha \beta}^X\right) \Bigg] \\
 &+\Delta q_a^X \sum_{(c \neq a) \in X} \Delta q_c^X \frac{\partial \gamma_{a c}}{\partial r_a}
 +\sum_{(b \neq a) \in X} \frac{\partial E_{a b}^{\mathrm{rep}}}{\partial r_a}
\end{align*}
 is identical to the DFTB gradient with the exception of the internal Lagrangian term
\begin{equation}
    W_{\mu \nu}^{\prime X}=\frac{1}{2} \sum_{\rho \sigma} P_{\mu \rho}^X H_{\rho \sigma}^X P_{\sigma \nu}^X,-\frac{1}{2} \sum_{\rho \sigma} P_{\mu \rho}^X V_{\rho \sigma}^X P_{\sigma \nu}^X,
\end{equation}
 which includes the electrostatic interaction with all other fragments ($V_{\rho \sigma}^X$).

The calculation of the gradient of the embedding energy differentiates between two cases: i) If atom $\alpha$ is part of the pair $IJ$, the gradient yields
\begin{align*}
\frac{\partial \Delta E_{I J}^{em}}{\partial r_{\alpha x}} &= \Delta \Delta q_\alpha^{I J} \sum_{K \neq I, J}^N \sum_{C \in K} \Delta q_C^K \frac{\partial \gamma_{\alpha C}}{\partial r_{\alpha x}} \\
& +\sum_{\mu \in \alpha} \sum_\nu\left(\Delta \tilde{W}_{\mu \nu}^{I J, \alpha} S_{\mu \nu}^{I J}+\Delta P_{\mu \nu}^{I J} \frac{\partial S_{\mu \nu}^{I J}}{\partial r_{\alpha x}}\right) \\
&\times \sum_{K \neq I, J}^N \sum_{C \in K} \gamma_{\alpha C} \Delta q_C^K,
\numberthis
\end{align*}
where 
\begin{equation}
    \Delta \tilde{W}_{\mu \nu}^{I J, \alpha}=\tilde{W}_{\mu \nu}^{I J, \alpha}-\left(\tilde{W}_{\mu \nu}^{I, \alpha} \oplus \tilde{W}_{\mu \nu}^{J, \alpha}\right)
\end{equation}
and
\begin{equation}
    \Delta P_{\mu \nu}^{I J}=P_{\mu \nu}^{I J}-\left(P_{\mu \nu}^I \oplus P_{\mu \nu}^J\right).
\end{equation}
ii) If atom $\alpha$ is not part of the pair $IJ$, the embedding energy gradient yields
\begin{align}
\begin{split}
\frac{\partial \Delta E_{I J}^{em}}{\partial r_{\alpha x}}&= \Delta q_\alpha^K \sum_{A \in I J} \Delta \Delta q_A^{I J} \frac{\partial \gamma_{A \alpha}}{\partial r_{\alpha x}}  \\
& +\sum_{\mu \in \alpha} \sum_\nu\left(\tilde{W}_{\mu \nu}^{K, \alpha} S_{\mu \nu}^K+P_{\mu \nu}^K \frac{\partial S_{\mu \nu}^K}{\partial r_{\alpha x}}\right) \\
&\times \sum_{A \in I J} \gamma_{A \alpha} \Delta \Delta q_A^{I J}, 
\end{split}
\end{align}
where
\begin{equation}
     \tilde{W}_{\mu \nu}^{X, \alpha}=-\frac{1}{2} \sum_{\rho \sigma} P_{\mu \rho}^X \frac{\partial S_{\rho \sigma}^X}{\partial r_{\alpha x}} P_{\sigma \nu}^X.
\end{equation}
The gradient of the ES-DIM energy also requires two different expressions: i) For atom $\alpha$ on fragment $I$ the gradient is
\begin{align}
\begin{split}
\frac{\partial E_{I J}}{\partial r_{\alpha x}}&=\Delta q_\alpha^I \sum_{B \in J} \Delta q_B^J \frac{\partial \gamma_{\alpha B}}{\partial r_{\alpha x}}\\
&+\sum_{\mu \in \alpha} \sum_\nu\left(\tilde{W}_{\mu \nu}^{I, \alpha} S_{\mu \nu}^I+P_{\mu \nu}^I \frac{\partial S_{\mu \nu}^I}{\partial r_{\alpha x}}\right)\\
&\times \sum_{B \in J} \Delta q_B^J \gamma_{\alpha B}.
\end{split}
\end{align}
ii) 
The gradient expression for atom $\alpha$ on fragment $J$ yields
\begin{align}
\begin{split}
\frac{\partial E_{I J}}{\partial r_{\alpha x}}&= \Delta q_\alpha^J \sum_{A \in I} \Delta q_A^I \frac{\partial \gamma_{A \alpha}}{\partial r_{\alpha x}}\\&
+\sum_{\mu \in \alpha} \sum_\nu\left(\tilde{W}_{\mu \nu}^{J, \alpha} S_{\mu \nu}^J+P_{\mu \nu}^J \frac{\partial S_{\mu \nu}^J}{\partial r_{\alpha x}}\right)\\
&\times \sum_{A \in I} \Delta q_A^I \gamma_{A \alpha}.
\end{split}
\end{align}
Using the previous expressions, the ground-state FMO-DFTB gradient is then calculated according to Eq. \ref{eq:fmo_gs_grad}.

\subsection{Excited states within FMO-LC-TDDFTB}
As we already described the excited-state formalism of the FMO-LC-TDDFTB method in our previous work \cite{einsele_long-range_2023}, we will only give a short summary of the working equations in the following section.

The excited state wavefunction of the complete system in the framework of the FMO-DFTB method can be formulated as a linear combination of basis states of the monomer and pair fragments. The basis is constructed from singlet locally excited states on single fragments and charge-transfer states between two fragments. Thus, the total electronic excited state wavefunction is defined as  
\begin{equation}
\left|\Psi\right\rangle=\sum_{I}^N \sum_{m}^{N_{\mathrm{LE}}} c_I^m \left|\mathrm{LE}_I^m \right\rangle+ \sum_{I}^{N} \sum_{J \neq I}^{N} \sum_{m}^{N_{\mathrm{CT}}} c_{I\rightarrow J}^m \left| \mathrm{CT}_{I \to J}^{m}\right\rangle,
\end{equation}
where the coefficients of the LE and CT configuration state functions are obtained by solving the eigenvalue problem $\mathbf{Hc} = \mathbf{Ec}$.
The LE states 
\begin{equation}
    \left|\mathrm{LE}_{\mathrm{I}}^m\right\rangle = \sum_{i \in I}\sum_{a \in I}  X_{ia}^{m(I)} | \Phi_{I}^{i\to a}\rangle
    \label{eq:LE_states}
\end{equation}
are obtained from the excited-state calculation of the monomer fragments, where $X_{ia}^{m(I)}$ is the one-particle transition density matrix of the $m$-th excited state of fragment $I$ in the monomer MO basis and $|\Phi_{I}^{i\to a}\rangle$ is the singlet configuration state function of the excitation from the occupied orbital i to the virtual orbital a on fragment $I$.

The charge-transfer states of a fragment pair including monomer $I$ and monomer $J$ can be expressed as
\begin{align}
    \left|\mathrm{CT}_{I \to J}^{m}\right\rangle = \sum_{i \in I}\sum_{a \in J} X_{ia}^{m(I \to J)} | \Phi_{I \to J}^{i\to a} \rangle,
    \label{eq:CT_states}
\end{align}
where the transition between the fragments is restricted to the occupied orbitals of monomer $I$ and the virtual orbitals of monomer $J$. 

The excited-state energies of the LEs and CTs are calculated using the linear-response formalism of the LC-DFTB method.\cite{humeniuk_long-range_2015} The excited states can be obtained by solving the non-Hermitian eigenvalue equation
\begin{equation}
    \left(\begin{array}{ll}
\mathbf{A} & \mathbf{B} \\
\mathbf{B} & \mathbf{A}
\end{array}\right)\left(\begin{array}{l}
\mathbf{X} \\
\mathbf{Y}
\end{array}
\right)=\mathbf{\Omega}\left(\begin{array}{cc}
\mathbf{1} & 0 \\
0 & -\mathbf{1}
\end{array}\right)\left(\begin{array}{l}
\mathbf{X} \\
\mathbf{Y}
\end{array}\right),\label{eq:Casida}
\end{equation}
where the matrices $\mathbf{A}$ and $\mathbf{B}$ are defined as
\begin{align}
\begin{split}
    A_{i a, j b} &=\delta_{i j} \delta_{a b}\left(\epsilon_a-\epsilon_i\right)+2 \sum_{A}\sum_{B} q^{ia}_A \gamma_{AB} q_B^{jb} \\
    &- \sum_{A} \sum_{B} q^{ij}_A \gamma_{AB}^{\mathrm{lr}} q^{ab}_B
\end{split}
\\
    B_{i a, j b} &= 2 \sum_{A}\sum_{B} q^{ia}_A \gamma_{AB} q_B^{jb} - \sum_{A} \sum_{B} q^{ib}_A \gamma_{AB}^{\mathrm{lr}} q^{aj}_B
\end{align}
in the framework of the LC-DFTB method. The atomic transition charges between MOs are expressed as
\begin{equation}
q_{A}^{i j}=\frac{1}{2} \sum_{\mu \in A} \sum_{\nu}\left(C_{\mu i} C_{\nu j}+C_{\nu i}C_{\mu j}\right) S_{\mu \nu}.
\label{eq:q_between_mos}
\end{equation}
Alternatively, if the Tamm-Dancoff (TDA) approximation is employed in the excited-state calculations, Eq. \ref{eq:Casida} is reduced to the Hermitian eigenvalue problem
\begin{equation}
    \mathbf{A} \mathbf{X} = \mathbf{\Omega} \mathbf{X},
\end{equation}
which is solved by the Davidson algorithm to obtain the lowest eigenvalues \cite{davidson_iterative_1975}.
\begin{figure}[!t]
    \centering
    \includegraphics[width=\linewidth]{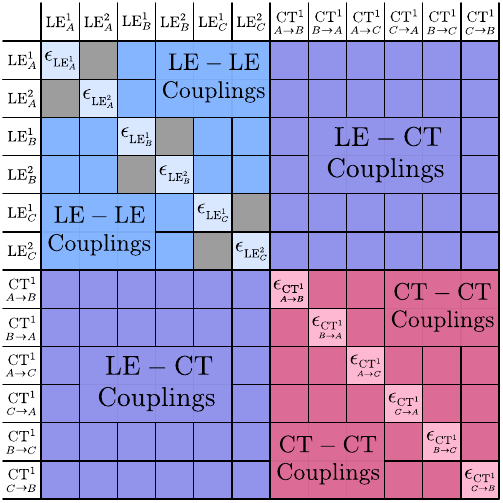}
    \caption{Illustration of the excited-state Hamiltonian of the FMO-LC-TDDFTB method. The Hamiltonian is shown for an example trimer system with two LE states for each monomer and one CT state for each pair.}
    \label{fig:excited-state-hamiltonian}
\end{figure}

To obtain the excited states of the complete system in the framework of the FMO-LC-TDDFTB method, the complete excited-state Hamiltonian $\mathbf{H^{Exc-FMO}}$, which is depicted in Fig. \ref{fig:excited-state-hamiltonian}, must be diagonalized. The construction of the Hamiltonian requires the couplings between the locally excited and charge-transfer basis states. The diagonal matrix elements of the Hamiltonian are represented by the energies of the respective basis states, obtained from the TDA-DFTB calculations. As derived in Refs. \citep{fujita_development_2018,einsele_long-range_2023}, the off-diagonal matrix elements can be split in one-electron 
\begin{equation}\label{One_electron_hamiltonian}
    \langle \Phi_{I \to J}^{i\to a}|\mathrm{H}_{1e}|\Phi_{K \to L}^{j \to b} \rangle = \delta_{I K} \delta_{i j} H_{a b}^{\prime}-\delta_{J L } \delta_{a b} H_{i j}^{\prime}
\end{equation}
and two-electron contributions
\begin{align}\label{two_electron_hamiltonian}
\begin{split}
    \langle \Phi_{I \to J}^{i\to a} | \mathrm{H}_{2 e} |\Phi_{K \to L}^{j \to b} \rangle &=2\left(i^{(I)} a^{(J)} \mid j^{(K)} b^{(L)}\right)\\ 
    &-\left(i^{(I)} j^{(K)} \mid a^{(J)} b^{(L)}\right).
\end{split}
\end{align}
The one-electron part vanishes for the coupling between the LE states according to the Slater-Condon rules and thus, the LE-LE coupling in the tight-binding formalism is given by
\begin{align*}\label{le_le_coupling}
\left\langle \mathrm{LE}_{I}^{m}\right|\mathrm{H}& \left| \mathrm{LE}_{J }^{n}\right\rangle 
= 2 \sum_{A\in I} \sum_{B \in J} q_{\mathrm{tr}, A}^{m(I)} \gamma_{AB} q_{\mathrm{tr}, B}^{n(J)} \numberthis \\
-& \sum_{A\in IJ} \sum_{B \in IJ} 
\sum_{ia \in I} \sum_{jb \in J} X_{ia}^{m(I)} X_{jb}^{n(J)} q_{A}^{ij} \gamma_{AB}^{\mathrm{lr}} q_{B}^{ab},
\end{align*}
where
\begin{equation}
    q_{\mathrm{tr}, A}^{m(I)} = \sum_{ia} q_{A}^{ia} X_{ia}^{m(I)} 
\end{equation}
is the transition charge of the $m$-th excited state of the fragment $I$ on atom $A$.
The first term represents the Coulomb interaction between the transition densities of fragments $I$ and $J$, whereas the second terms represents the exchange interaction. If the fragment pair is classified as a far separated pair, the ES-DIM approximation is used and the exchange interaction is neglected.

The coupling of the LE state on fragment $I$ and the charge-transfer state between fragment $J$ and $K$ is defined as
\begin{align*}\label{le_ct_coupling}
    \left\langle\mathrm{\mathrm{LE}}_{I}^{m}\right|\mathrm{H}& \left| \mathrm{CT}_{J \to K}^{n}\right\rangle = \\& \delta_{I J} \sum_{ia \in I}\sum_{b \in K} X_{ia}^{m(I)} X_{ib}^{n(I \to K)} H_{a b}^{\prime} \\ &-\delta_{I K}
    \sum_{ia \in I}\sum_{j \in J} X_{ia}^{m(I)}X_{ja}^{n(J \to I)} H_{i j}^{\prime} \\ 
+& 2\sum_{A \in I}\sum_{B \in JK} q_{\mathrm{tr}, A}^{m(I)} \gamma_{AB} q_{\mathrm{tr}, B}^{n(J \to K)} \numberthis \\
- \sum_{A \in IJ}\sum_{B \in IK
} &\sum_{ia \in I}  \sum_{j \in J}\sum_{b \in K}X_{ia}^{m(I)} X_{jb}^{n(J \to K)} q_{A}^{ij} \gamma_{AB}^{\mathrm{lr}} q_{B}^{ab},  
\end{align*}
where $H_{a b}^{\prime}$ are matrix elements of the orthogonalized Hamiltonian of the full system, which is given by
\begin{equation}
    \mathbf{H}^{\prime} = \mathbf{S}^{-\frac{1}{2}} \mathbf{H}^{LCMO} \mathbf{S}^{-\frac{1}{2}}.
    \label{eq:orthogonalized_hamiltonian}
\end{equation}
Here, $\mathbf{S}$ is the overlap matrix of the complete system and the non-orthogonalized Hamiltonian is constructed from the Hamiltonian matrices of the monomer and pair fragments as
\begin{equation}
    \mathbf{H}^{LCMO} = \bigoplus_{I} \mathbf{H}^{I} + \bigoplus_{I>J}(\mathbf{H}^{IJ} - \mathbf{H}^{I} \oplus \mathbf{H}^{J}).
\end{equation}
Compared to the LE-LE coupling, the one-electron contribution to the LE-CT coupling does not vanish. However, if the LE and CT state do not share a monomer and the fragments of the basis states are not in close proximity to each other, the contributions get zero as the matrix elements of the orthogonalized Hamiltonian $\mathbf{H}^{'}$ (\textit{cf}. Eq. \ref{eq:orthogonalized_hamiltonian}) are zero for ES-DIM pair fragments. In the case of far separated fragments, the ES-DIM approximation is used and the only remaining part of the LE-CT coupling is the Coulomb interaction
\begin{equation}
    \left\langle\mathrm{\mathrm{LE}}_{I}^{m}\left|\mathrm{H}\right| \mathrm{CT}_{J \to K}^{n}\right\rangle = 2\sum_{A \in I}\sum_{B \in JK} q_{\mathrm{tr}, A}^{m(I)} \gamma_{AB} q_{\mathrm{tr}, B}^{n(J \to K)}.
\end{equation}

The one-electron contributions of the CT-CT coupling are included in the diagonal matrix elements obtained from the LC-TD-DFTB calculations. The remaining two-electron contribution of the coupling between two charge-transfer states is defined as
\begin{align}\label{ct_ct_coupling}
\begin{split}
\left\langle \mathrm{CT}_{I \to J}^{m} \right|& \mathrm{H}\left| \mathrm{CT}_{K \to L}^{n}\right\rangle = \\ &2 \sum_{A\in IJ} \sum_{B \in KL} q_{\mathrm{tr}, A}^{m(I \to J)} \gamma_{AB} q_{\mathrm{tr}, B}^{n(K \to L)} \\
&- \sum_{i \in I}\sum_{a \in J}\sum_{j \in K}\sum_{b \in L} \sum_{A \in IK} \sum_{B \in JL} X_{ia}^{m(I \to J)}\\
&\times X_{jb}^{n(K \to L)}q_{A}^{ij} \gamma_{AB}^{\mathrm{lr}} q_{B}^{ab}. 
\end{split}
\end{align}
If the fragments $I$ and $K$ or $J$ and $L$ are not in close spatial proximity, the ES-DIM approximation is utilized and thus, the exchange contribution of the CT-CT coupling is neglected.
\newline
\textbf{Excited state gradients}

In the following, we derive the gradient expressions for the excited-state energies of the locally excited and charge-transfer basis states. However, we will only give a short summary of the derivation, a more detailed description is available in Refs. \cite{heringer_analytical_2007,humeniuk_dftbaby_2017}.

Utilizing Furche's variational TD-DFT formalism \cite{furche_adiabatic_2002}, excitation energies of TDA-DFTB are stationary points of the functional
\begin{equation}
    G[\mathbf{X},\mathbf{\Omega}, \mathbf{C}]=\mathbf{X^T}\mathbf{A}\mathbf{X}-\mathbf{\Omega}\left(\mathbf{X}\mathbf{X}^T-1\right),
\end{equation}
which is part of the auxiliary functional 
\begin{align}\label{eq:gradient_lagrangian}
\begin{split}
L\left[\textbf{X}, \mathbf{\Omega}, \textbf{C}, \textbf{Z}, \textbf{W}\right] &=\mathbf{X^T}\mathbf{A}\mathbf{X}-\mathbf{\Omega}\left(\mathbf{X}\mathbf{X}^T-1\right)\\
&+\sum_{i, a} Z_{i a} H_{i a}\\
&-\sum_{p, q, p \leq q} W_{p q}\left(S_{p q}-\delta_{p q}\right),
\end{split}
\end{align}
where $\mathbf{W}$ and $\mathbf{Z}$ are Lagrange multipliers. The minimization of the auxiliary functional leads to a set of conditions, i.a. the stationary conditions of the MO coefficients
\begin{equation}
    \frac{\partial L}{\partial \mathbf{C}} = 0,
    \label{eq:dL-dc-condition}
\end{equation}
which can be utilized to determine the Lagrange multipliers $\mathbf{W}$ and $\mathbf{Z}$ (\textit{cf}. Appendix \ref{appendix:langrange_multipliers}).
Subsequently, the gradient of the excitation energy can be formulated after some lengthy mathematical derivations as
\begin{equation}
    \frac{d \Omega}{d R} =\frac{\partial G}{\partial R}+\sum_{i a} Z_{i a} \frac{\partial H_{i a}}{\partial R}-\sum_{p, q, p \leq q} W_{p q} \frac{\partial S_{p q}}{\partial R},
\end{equation}
where 
\begin{align}
\frac{\partial G}{\partial R}&= \sum_{i a, j b} X_{i a}\frac{\partial A_{ij,a b}}{\partial R}X_{j b} \\
\begin{split}
 &= \sum_{i a, j b} X_{i a}\left\{\delta_{i j} \frac{\partial H_{a b}}{\partial R}-\delta_{a b} \frac{\partial H_{i j}}{\partial R} \right. \\
 & \left. +2 \frac{\partial(i a \mid j b)}{\partial R}-\frac{\partial(i j \mid a b)_{\mathrm{lr}}}{\partial R}\right\}X_{j b}
\end{split}
 \\ 
\begin{split}
&=\sum_{a, b} \frac{\partial H_{a b}}{\partial R} \sum_iX_{i a}X_{i b}
-\sum_{i, j} \frac{\partial H_{i j}}{\partial R} \sum_aX_{i a}X_{j a} \\ & +2 \sum_{i a, j b} \frac{\partial(i a \mid j b)}{\partial R}X_{i a}X_{j b}
-\sum_{i a, j b} \frac{\partial(i j \mid a b)_{\operatorname{lr}}}{\partial R}X_{i a}X_{j b}.
\end{split}
\end{align}
After transformation from the MO basis to the AO basis, the expression of the gradient of the excitation energy is given by
% \begin{align}
%     T_{ab} &= \sum_i X_{ia}X_{ib} \\
%     T_{ij} &= \sum_a X_{ia}X_{ja}
% \end{align}
\begin{align}
\begin{split}
    \frac{d \Omega}{d R} &=\sum_{\alpha, \beta} \frac{\partial H_{\alpha \beta}}{\partial R}\left\{T_{\alpha \beta}^{\mathrm{v}-\mathrm{v}}-T_{\alpha \beta}^{\mathrm{o}-\mathrm{o}}+Z_{\alpha \beta}\right\}\\
    &-\sum_{\alpha, \beta} \frac{\partial S_{\alpha \beta}}{\partial R} W_{\alpha \beta} \\ 
    &+2 \sum_{\alpha, \beta} X_{\alpha \beta} \sum_{\gamma, \delta} \frac{\partial(\alpha \beta \mid \gamma \delta)}{\partial R} X_{\gamma \delta}\\
    &-\sum_{\alpha, \beta} X_{\alpha \beta} \sum_{\gamma, \delta} \frac{\partial(\alpha \gamma \mid \beta \delta)_{1 \mathrm{r}}}{\partial R}X_{\gamma \delta},
\end{split}
\end{align}
where 
\begin{align}
    T_{\alpha \beta}^{\mathrm{v}-\mathrm{v}} &= \sum_a \sum_b C_{\alpha a} C_{\beta b}\sum_i X_{ia}X_{ib} \\
    T_{\alpha \beta}^{\mathrm{o}-\mathrm{o}} &= \sum_i \sum_j C_{\alpha i} C_{\beta j}\sum_a X_{ia}X_{ja}.
\end{align}
The gradient of a locally excited basis state is equivalent to the LC-TD(A)-DFTB gradient and requires no further calculations. However, the calculation of the gradient of a charge-transfer state requires some preliminary transformations. As mentioned earlier, a CT state from fragment $I$ to fragment $J$ is restricted to the transition of the occupied orbitals of $I$ and the virtual orbitals of $J$. In order to calculate the gradient of the $m$-th CT state of $IJ$, the transition density matrix $\mathbf{X}^{m(I \to J)}$ in the MO basis between the occupied and virtual orbitals of the two fragments is transformed into the MO basis of the fragment pair $\mathbf{X}^{m(IJ)}$ by utilizing the overlap matrices between the monomer and pair orbitals $S_{\mu,\lambda}^{I,IJ}$ and $S_{\nu,\sigma}^{J,IJ}$: 
\begin{align}\label{eq:ct-eigenvector-transform}
\begin{split}
    X_{jb}^{m(IJ)} &= \sum_{\mu \in I}\sum_{\nu \in J}\sum_{\lambda \sigma \in IJ} \sum_{i \in I} \sum_{a \in J}C_{\mu i}^{I}S_{\mu,\lambda}^{I,IJ} C_{\lambda j}^{IJ} \\ 
    &\times X_{ia}^{m(I \to J)} C_{\nu a}^{J} S_{\nu,\sigma}^{J,IJ} C_{\sigma b}^{IJ}.
\end{split}
\end{align}
% where $S_{\mu,\lambda}^{I,IJ}$ and $S_{\nu,\sigma}^{J,IJ}$ are defined as
% \begin{equation}
%     S_{\mu,\lambda}^{I,IJ} = \left<\varphi_{\mu}^{I}|\varphi_{\lambda}^{IJ}\right>
% \end{equation}
Subsequently, the gradients of a charge-transfer state from fragment $I$ to $J$ are obtained from the LC-TD(A)-DFTB gradient calculation of the pair $IJ$ with the excited state eigenvector obtained according to Eq. \ref{eq:ct-eigenvector-transform}.
\subsection{Nonadiabatic excited-state Ehrenfest dynamics}
In this section, we present the theoretical methodology of nonadiabatic excited-state molecular dynamics in the framework of the quasi-diabatic basis states of our FMO-LC-TDDFTB approach. 

In our semiclassical approach, the dynamics of the atomic nuclei are governed by Newton's equation, whereas the dynamics of the electronic motion is treated quantum mechanically. Thus, the electronic excited state wavefunction is parametrically dependent on the nuclear coordinates $\mathbf{R}(t)$ and can be expanded in the basis of the quasi-diabatic LE and CT states of the molecular aggregates as follows
\begin{align*}\label{Basis_expansion:le_ct}
\left|\Psi(\mathbf{r;R}(t))\right\rangle&=\sum_{I}^N \sum_{m}^{N_{\mathrm{LE}}} c_I^m(t) \left|\mathrm{LE}_I^m(\mathbf{r;R}(t)) \right\rangle \numberthis \\
&+ \sum_{I}^{N} \sum_{J \neq I}^{N} \sum_{m}^{N_{\mathrm{CT}}} c_{I\rightarrow J}^m(t) \left| \mathrm{CT}_{I \to J}^{m}(\mathbf{r;R}(t))\right\rangle,
\end{align*}
where $c_I^m(t)$ and $c_{I\rightarrow J}^m(t)$ are the time-dependent expansion coefficients for the LE and CT states. Employing the notation $\left|\psi_n\right\rangle$ for a quasi-diabatic basis state, Eq. (\ref{Basis_expansion:le_ct}) becomes
\begin{equation}\label{Basis_expansion:basis_state}
\left|\Psi(\mathbf{r;R}(t))\right\rangle=\sum_{n} c_n(t) \left|\psi_n(\mathbf{r;R}(t)) \right\rangle.
\end{equation}
After inserting Eq. (\ref{Basis_expansion:basis_state}) into the time-dependent Schrödinger equation, a set of coupled differential equations for the expansion coefficients is obtained in the diabatic representation
\begin{align}
\begin{split}
    \mathrm{i}\hbar\frac{d c_n(t)}{d t}&=\sum_m c_m(t)\left[H_{nm}^{Exc-FMO}(\mathbf{R}(t)) \right. \\
    & \left. -\mathrm{i}\hbar D_{nm}\left(\mathbf{R}(t)\right) \right],
\end{split}
\end{align}
where $H_{nm}^{Exc-FMO}$ is a matrix element of the excited state Hamiltonian of the FMO-LC-TDDFTB method (\textit{cf}. Fig. \ref{fig:excited-state-hamiltonian}) and $D_{nm}$ is the nonadiabatic coupling between two quasi-diabatic states. Due to the fact that the locally excited and charge-transfer states are adiabatic states of the single fragments, the calculation of the nonadiabatic coupling between the adiabatic states on a fragment is required. Thus, the nonadiabatic coupling
\begin{equation}
     D_{nm}(\mathbf{R}(t)) =
        \left\langle \psi_n(\mathbf{r;R}(t))  \Big| \frac{\partial \psi_m(\mathbf{r;R}(t))}{\partial t} \right\rangle
\end{equation}
 is calculated between LE states on the same monomer and between CT states on the same pair fragment, while it is zero between all other quasi-diabatic basis states. In our simulations, the nonadiabatic coupling is obtained from the scalar product of the nonadiabatic coupling vector and the nuclear velocities
 \begin{equation}
     D_{nm}(\mathbf{R}(t)) = \left\langle \psi_n(\mathbf{r;R}(t))  \right| \nabla_{\mathbf{R}} \left| \psi_m(\mathbf{r;R}(t))  \right\rangle \cdot \Dot{\mathbf{R}}(t).
 \end{equation}
As the derivation of the nonadiabatic coupling vectors in the framework of the TD-LC-DFTB method was recently published by Niehaus~\cite{niehaus_exact_2023}, we will only briefly present the working equations for the TDA-LC-DFTB approach. The nonadiabtic coupling between two adiabatic states on the same fragment is given by
\begin{align}
\begin{split}
    \left\langle \psi_n  \right| \nabla_{\mathbf{R}} \left| \psi_m  \right\rangle &=\frac{1}{\Omega_n - \Omega_m}\bigg[ \sum_{\alpha, \beta} \frac{\partial H_{\alpha \beta}}{\partial R} \\& \times 
    \left\{\Tilde{T}_{\alpha \beta}^{\mathrm{v}-\mathrm{v},nm}-\Tilde{T}_{\alpha \beta}^{\mathrm{o}-\mathrm{o}, nm}+Z_{\alpha \beta}^{nm}\right\}\\ 
    &-\sum_{\alpha, \beta} \frac{\partial S_{\alpha \beta}}{\partial R} W_{\alpha \beta}^{nm}  \\ 
    &+2 \sum_{\alpha, \beta} X_{\alpha \beta}^{n} \sum_{\gamma, \delta} \frac{\partial(\alpha \beta \mid \gamma \delta)}{\partial R} X_{\gamma \delta}^{m} \\
    &-\sum_{\alpha, \beta} X_{\alpha \beta}^{n} \sum_{\gamma, \delta} \frac{\partial(\alpha \gamma \mid \beta \delta)_{1 \mathrm{r}}}{\partial R}X_{\gamma \delta}^{m}\bigg],
\end{split}
\end{align}
where 
\begin{align}
    \Tilde{T}_{\alpha \beta}^{\mathrm{v}-\mathrm{v},nm} &= \sum_a \sum_b C_{\alpha a} C_{\beta b}\sum_i \frac{1}{2}\left(X_{ia}^{n}X_{ib}^{m}+X_{ia}^{m}X_{ib}^{n} \right) \\
    \Tilde{T}_{\alpha \beta}^{\mathrm{o}-\mathrm{o},nm} &= \sum_i \sum_j C_{\alpha i} C_{\beta j}\sum_a \frac{1}{2} \left( X_{ia}^{n}X_{ja}^{m} + X_{ia}^{m}X_{ja}^{n} \right).
\end{align}
The Lagrange multipliers $\mathbf{Z^{nm}}$ and $\mathbf{W^{nm}}$ for the nonadiabatic couplings can be derived in analogy to the TDA-LC-DFTB gradient \cite{wu_nonadiabatic_2022,niehaus_exact_2023}. 
As we are using Ehrenfest dynamics in the quasi-diabatic basis of LE and CT states, we approximate the excited-state forces as the linear combination of the TDA-LC-DFTB gradients of all basis states
\begin{equation}
    F_{R} = - \sum_{n} \left|c_n\right|^2 \nabla_{R} E_n,
\end{equation}
where $c_n$ are the expansion coefficients of the quasi-diabatic basis states.

\subsection{Computational Details}\label{sec:comp_details}
In the present section, the computational technicalities of the quantum mechanical simulations will be described.

We employed the ob2\cite{vuong_parametrization_2018} parameter set for the calculations of the anthracene system and the mio\cite{elstner_self-consistent-charge_1998,niehaus_application_2001} parameter set for the molecular dynamics of the BTBT aggregates. In all simulations in this work, a long-range radius of 3.03 $a_0$ was used for the LC-DFTB approach.

The Mercury\cite{macrae_mercury_2020} program was utilized to generate the molecular structures of the anthracene and benzene aggregates from their crystal structures\cite{mason_crystallography_1964,cox_crystal_1997}.

In case of the molecular dynamics simulations of the anthracene system, the DFT-D3\cite{grimme_consistent_2010,grimme_effect_2011} dispersion correction was employed in conjunction with the ob2 dispersion parameters.
To prevent the molecular aggregates from moving apart during the course of the molecular dynamics simulations, the gradient of the harmonic potential
\begin{equation}
    V(\mathbf{R}) =\frac{1}{2}k (\mathbf{R} - \mathbf{R_{0}})^2
\end{equation}
was added to the total gradient of the system, where $\mathbf{R_0}$ are the initial coordinates of the molecular system and the force constant $k$ is $1.0 \cdot 10^{-6}$$ E_h/a_{0}^2$

The initial velocities of the molecular dynamics simulations were sampled from a Maxwell-Boltzmann distribution at a temperature of 300~K.
The classical dynamics of the nuclei was performed using the Velocity-Verlet algorithm with a time step of 0.1~fs.

For the molecular dynamics of the anthracene system (30 monomers), 50 trajectories were propagated for 20000 steps (2000 fs), respectively. In case of the excited-state dynamics of the model system of BTBT molecules (8 monomers), 10 trajectories were propagated for 200000 steps (20 ps). 
In case of both excited-state dynamics simulations, the number of CT states is one, whereas the number of LE states is set to three for the anthracene system and two for the BTBT aggregate.

The excited-state dynamics of our FMO-LC-TDDFTB method was implemented in our own software package DIALECT, which is available on Github\cite{noauthor_dialect_2024}.

%% file: results.tex
\section{Results and Discussion}\label{sec:results}
In this section, the accuracy of the excited-state gradients of the LE and CT states will be investigated by comparing the analytical results with the numerical values. Subsequently, a simulation of the excited-state molecular dynamics of a anthracene aggregate is presented to illustrate the efficacy of our method to describe exciton dynamics. Furthermore, we show the molecular dynamics simulation of the excited-state charge-transfer dynamics of the BTBT system to demonstrate the applicability of our method to charge-transfer dynamics of organic materials. In addition, the applicability of our method for simulating the recombination of a CT state is shown for the BTBT molecular system.
\subsection{Accuracy of the excited-state gradients}
In order to investigate the accuracy of the excited-state gradients of the locally excited and charge-transfer states, the analytical excited-state gradients are compared to the numerical gradients.
\begin{figure}[t!]
    \centering
    \includegraphics[width=\linewidth]{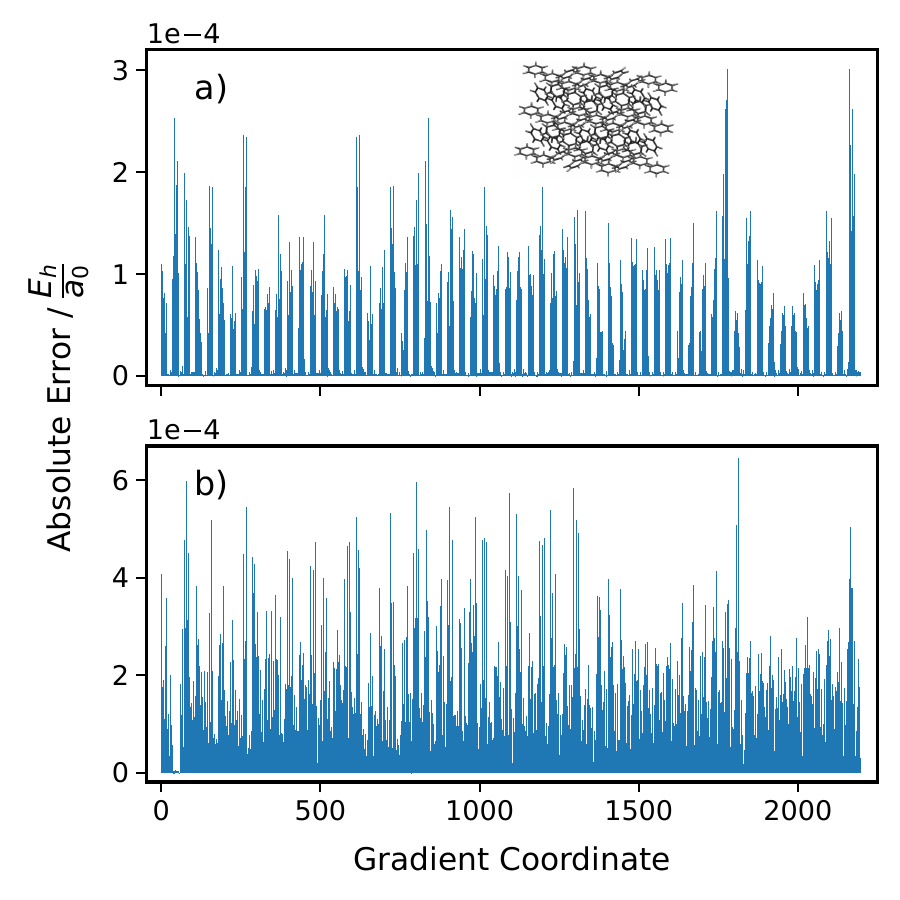}
    \caption{Absolute deviation of the analytical a) LE and b) CT gradients from the numerical gradients for the benzene crystal structure $(C_6H_6)_{61}$.}
    \label{fig:benezene_gradients}
\end{figure}
To this end, the excited-state gradients of a molecular aggregate of 61 benzene molecules, which was obtained from the crystal structure of benzene\cite{cox_crystal_1997}, are calculated. The deviations of the analytical gradient from the results of the numerical gradient are shown in Fig. \ref{fig:benezene_gradients}.
\begin{table}[b!]
\centering
\caption{Comparison of the analytical and numerical gradients for the benzene crystal.}
\label{tab:benzene_gradients}
\begin{tabular}{@{}c|cc@{}}
\toprule
   gradients & rmsd    & max error \\ \midrule
LE & 0.000058 & 0.00030    \\
CT & 0.00015  & 0.00064   
\end{tabular}
\end{table}
In case of the LE states, the gradients of the first excited-state of all benzene monomers are compared to the numerical results. The CT gradients are limited to the transitions from the first monomer fragment to the rest of the monomers. The root-mean-square deviation (rmsd) and the maximum deviation of the analytical gradient from the numerical results is shown in Table \ref{tab:benzene_gradients}.
While the LE gradients show deviations in the magnitude of $10^{-5} E_{\mathrm{h}}/a_0$ with maximum values reaching $3 \cdot 10^{-4} E_{\mathrm{h}}/a_0$, the errors of the CT gradients are in the range of $10^{-4} E_{\mathrm{h}}/a_0$ and reach a maximum of $6 \cdot 10^{-4} E_{\mathrm{h}}/a_0$. These deviations should be sufficiently small to accurately simulate the excited-state dynamics of molecular systems. 

\subsection{Excited-state exciton dynamics of anthracene}
\begin{figure}[t!]
    \centering
    \includegraphics[scale=1.0]{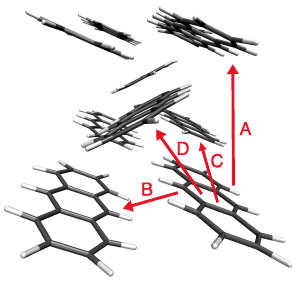}
    \caption{Illustration of the different excitonic couplings in the anthracene crystal structure.}
    \label{fig:anthracene_axes}
\end{figure}

\begin{figure}[t!]
    \centering
    \includegraphics[width=\linewidth]{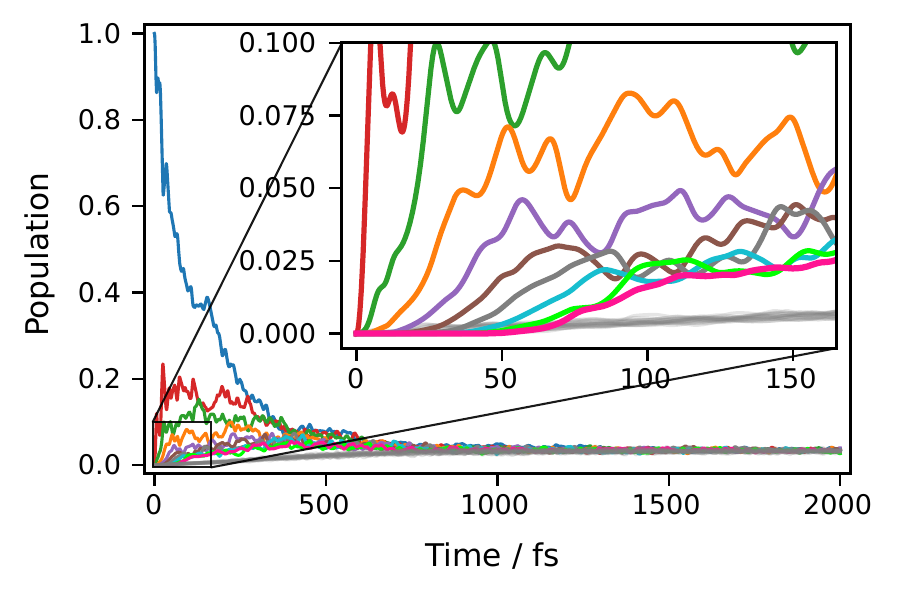}
    \caption{Population of the excited-states of the anthracene monomers averaged over all trajectories. The colored lines depict the populations of the anthracene monomers along the middle row of molecules, the gray lines display the populations of the monomers along the upper and lower horizontal rows.}
    \label{fig:averaged_populations}
\end{figure}

To investigate the excited-state dynamics of a molecular system of anthracene aggregates, a two-dimensional structure of 30 monomers (3 x 10) was cut from the crystal structure. A structure was chosen along the B-axis of the anthracene crystal that shows the highest excitonic coupling among the crystallographic axes of the anthracene system. The calculated excitonic coupling between the $S_1$ and $S_2$ states of the anthracene molecules for 4 crystallographic directions (cf. Fig. \ref{fig:anthracene_axes}) of the structure is shown in Tab. \ref{tab:coupling strengths}.
\begin{table}[b!]
\centering
\caption{Excitonic coupling strength (cf. Eq. \ref{le_le_coupling}) between the LE states of neighbouring molecules of the anthracene crystal along the axes in the crystal structure.}
\label{tab:coupling strengths}
\begin{tabular}{@{}c|cccc@{}}
\toprule
& \multicolumn{4}{c}{ abs. coupling strength / eV }\\
\toprule
LE state & A-axis & B-axis & C-axis & D-axis \\ \midrule
$S_1$ & 0.008 & 0.048 & 0.005 & 0.011 \\
$S_2$ & 0.003  & 0.013 & 0.009 & 0.013
\end{tabular}
\end{table}
\begin{figure}[b!]
    \centering
    \includegraphics[scale=1.0]{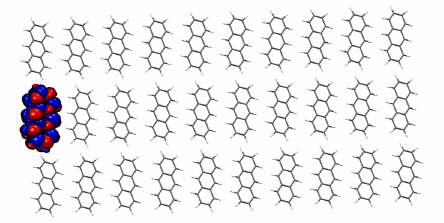}
    \caption{Initial state of the excited-state dynamics of the anthracene system.}
    \label{fig:initial_anthracene}
\end{figure}
The initial state of the excited-state dynamics is set to the first LE state on the outermost anthracene monomer of the middle row of the molecular system and is shown in Fig. \ref{fig:initial_anthracene}. As the simulation starts at the outermost monomer, a transfer of the exciton along the horizontal axis to the other side is expected. To observe this phenomenon, the excited-state population of each monomer averaged over all 50 trajectories is shown in Fig. \ref{fig:averaged_populations}. Here, the populations of the monomers along the middle horizontal row are displayed in bright colors, whereas the excited-state populations of the upper and lower row are shown in grey (cf. inset). As depicted in Fig. \ref{fig:averaged_populations}, the excited-state population of the initial state is gradually decreasing, while the excited-state population of the other monomers in the same row is increasing. 
At a simulation time of ca. 70--80 fs, the exciton transfer has progressed to the last monomer of the row and shows an increase of its populations.
\begin{figure}[b!]
    \centering
    \includegraphics[width=\linewidth]{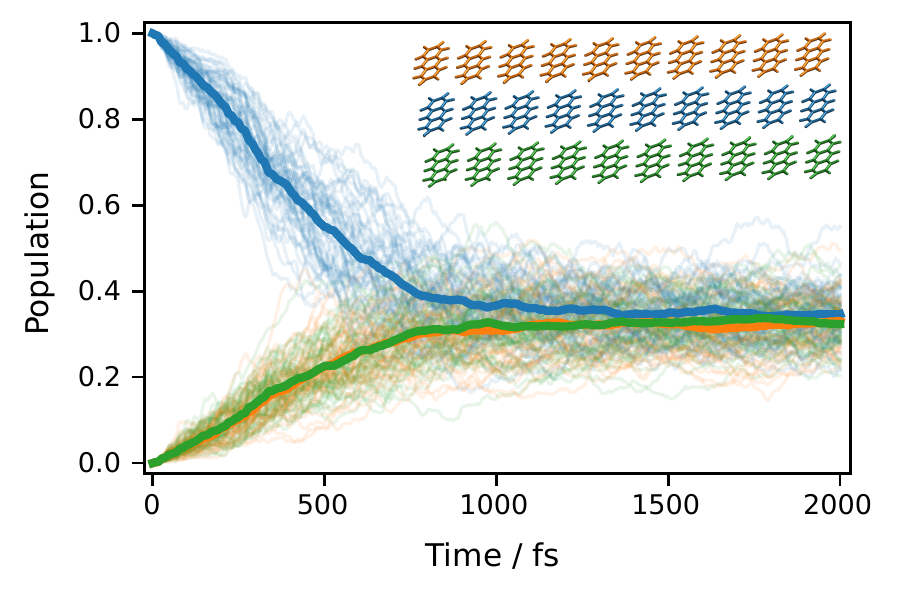}
    \caption{Excited-state populations along the b-axis of the anthracene crystal. The average population of all trajectories (thick lines) and the population of the individual trajectories (thin lines) are shown.}
    \label{fig:populations_axis}
\end{figure}
However, due to the low excitonic coupling between the different rows, the populations of the upper and lower anthracene monomers is only growing slowly. As the simulation time increases, the exciton is delocalized over the complete molecular structure of the anthracene aggregates.
\begin{figure*}[h!]
    \centering
    \includegraphics{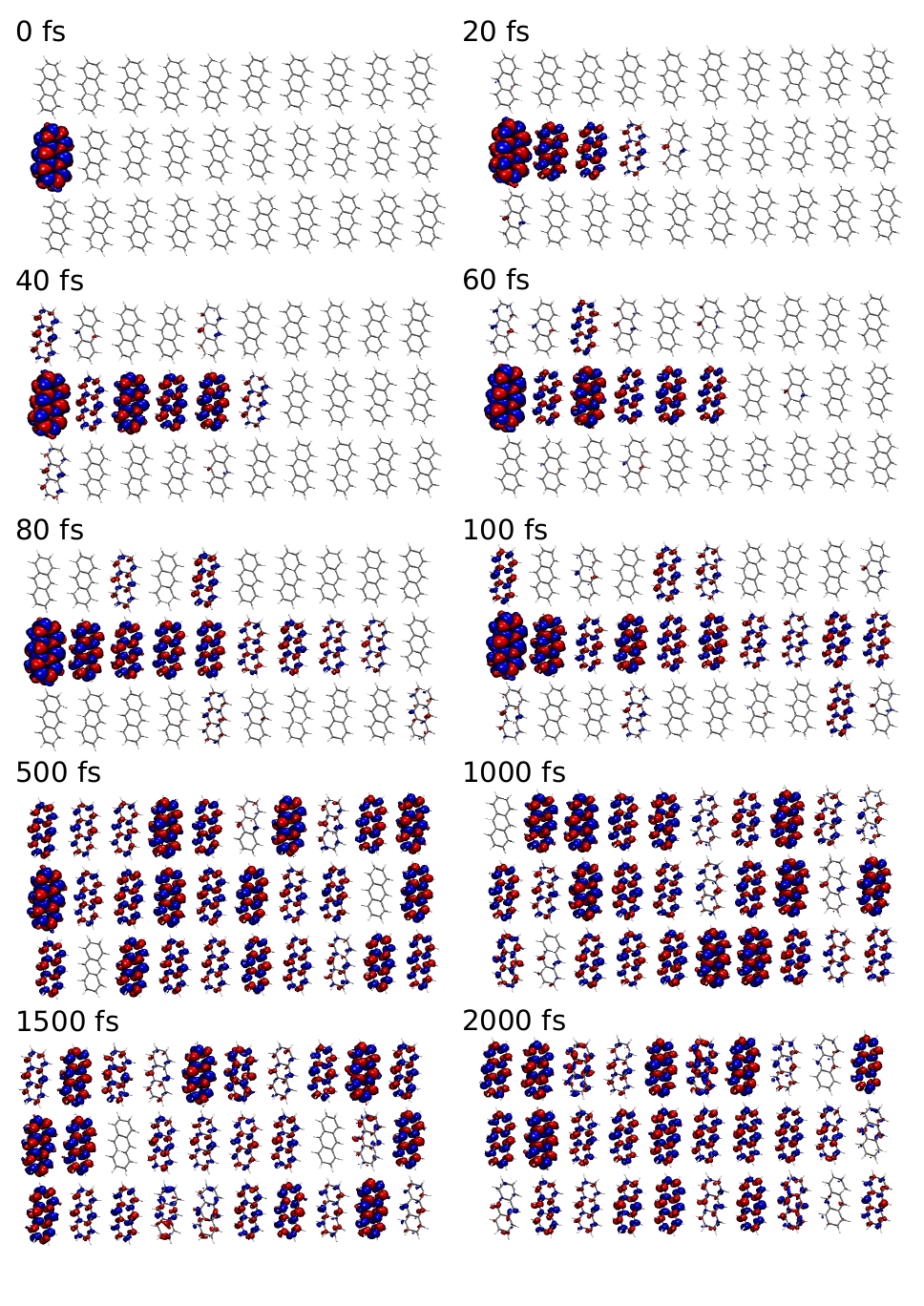}
    \caption{Transition densities of the excited-state configurations along a single example trajectory of the anthracene system.}
    \label{fig:tdms_along_trajectory}
\end{figure*}
From around 1500 fs onwards, the average excited-state population of all monomers shows only slight fluctuations and has reached a value of ca. 1/30, and thus, the exciton is equally delocalized over all monomers.
In order to illustrate this phenomenon in greater detail, the sum of the excited-state populations of the three horizontal rows of anthracene monomers is depicted in Fig. \ref{fig:populations_axis}. Up to a simulation time of 500~fs, the exciton dynamics is mainly taking place at the central row of anthracene molecules.

Between 500~fs and 1000~fs, the population of anthracene system is gradually delocalized over the whole aggregate. At around 1500~fs, the complete delocalization of the exciton is reached. However, compared to the average population of all trajectories, the excited-state populations of the individual trajectories show pronounced oscillations around the average population of $\frac{1}{3}$.

In the interest of a visual representation of the exciton dynamics, the transition density matrices for a single example trajectory are shown in Fig. \ref{fig:tdms_along_trajectory} at different simulation times. Between the start of the simulation and 100~fs, the initially localized population of the exciton is gradually transported along the central chain of anthracene molecules and populates the excited state of the last monomer. At the same time, a small part of the excited-state population is transfered to the upper and lower chain of anthracene monomers. Between 100~fs to 500~fs, the exciton population is partially transferred from the central anthracene chain to the upper and lower rows of anthracene molecules. Furthermore, a high degree of delocalization is already present at 500~fs. From this point of the simulation, the exciton population keeps fluctuating between the excited-states of all monomers of the system. However, the exciton is at no point of the simulation equally distributed between all monomers, instead showing varying degrees of delocalization on the three anthracene chains.

\subsection{Excited-state charge-transfer dynamics of BTBT}
In order to show the feasibility of the simulation of excited-state charge-transfer dynamics by employing our method, we investigate the excited-state dynamics of a model BTBT system, which is known as an organic field-effect transistor and a p-type semiconductor. 

\begin{figure}[t!]
    \centering
    \includegraphics[scale=1.0]{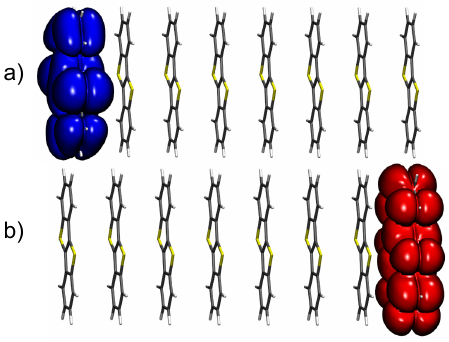}
    \caption{Initial charge-transfer state of the excited-state dynamics of the BTBT system. a) The hole (blue) and b) particle (red) density of the excited state is shown.}
    \label{fig:btbt_initial_state}
\end{figure}

To this end, a small molecular aggregate of 8 stacked BTBT monomers was prepared. The excited-state dynamics simulation starts from an initially far separated CT state, which is depicted in Fig. \ref{fig:btbt_initial_state}.

Due to the increased hole-conductive properties of the BTBT system, the movement of the hole should be favored compared to that of the particle. However, opposed to typical charge-transfer processes in solar cells or organic light-emitting diods, the movement of the CT states in our method is driven by the excitonic coupling between the excited-states and not by an external electric field. 
\begin{figure}[b!]
    \centering
    \includegraphics[width=\linewidth]{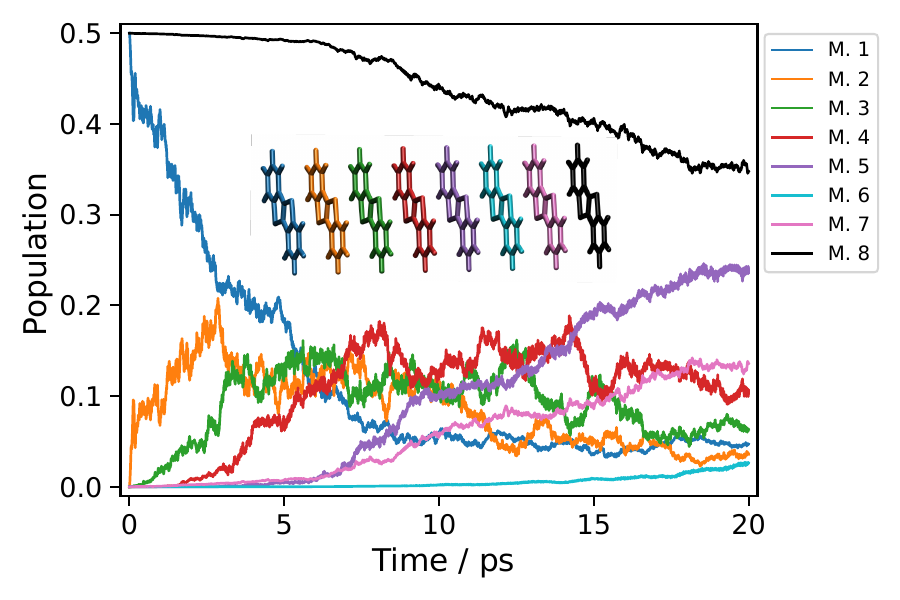}
    \caption{Mean excited-state populations of the BTBT monomers averaged for all trajectories.}
    \label{fig:btbt_average}
\end{figure}
The average excited-state population of the BTBT monomers for all 10 trajectories is shown in Fig. \ref{fig:btbt_average}. As the simulation starts from a CT state between the first and the last monomer of the BTBT chain, the populations of the respective states are initially 0.5. Compared to the population of the electron on the last monomer, which remains relatively constant for the first five picoseconds, the hole population on the first monomer shows a pronounced decline in conjunction with the increase of the excited-state populations on the second and third BTBT molecule. From 5~ps to 15~ps, the hole population is partially transferred from the first to the 5th monomer, showing a delocalization over the first five BTBT molecules.
Compared to the hole transfer process, the population of the electron is gradually transferred from the 8th to the 7th monomer, illustrating a much slower process.
\begin{figure*}[h!]
    \centering
    \includegraphics{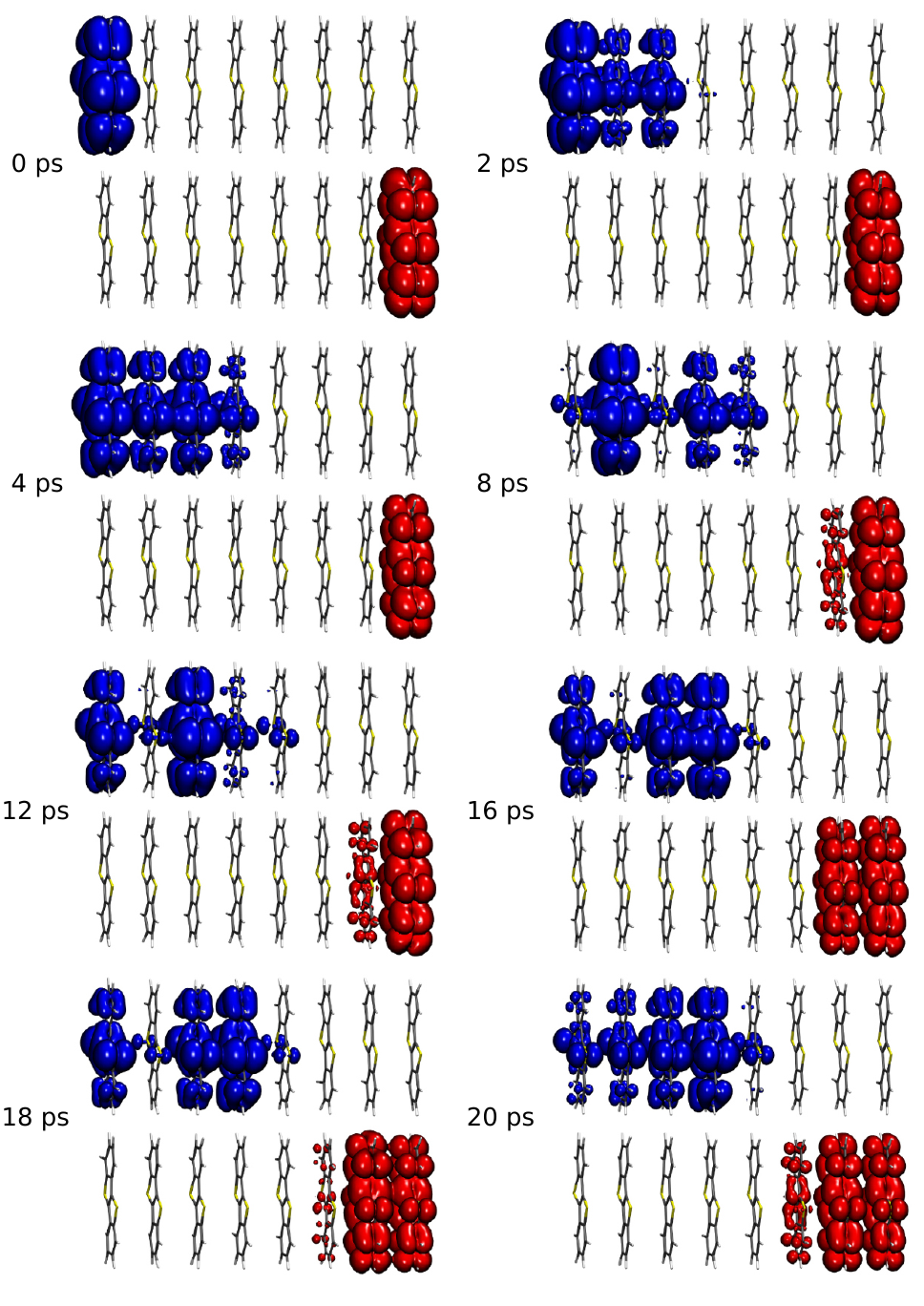}
    \caption{The hole (blue) and particle (red) densities of the excited states of the BTBT system of a single example trajectory.} 
    \label{fig:btbt_hole_particle}
\end{figure*}

At the end of the simulation, the majority of the population of the electron remains on the last BTBT molecules, showing a partial transfer to the 7th monomer and a marginal population of the CT contribution of the 6th monomer. Opposed to this, the population of the hole is predominantly located on the 5th BTBT molecule at the end of the simulation, showing a much higher mobility than the electron. However, the hole is still partially delocalized over the first four monomers, showing the least amount of population on the first and second molecule.

To obtain a more visual representation of the charge-transfer dynamics of the BTBT model system, the hole and particle densities for different simulation times of an example trajectory are displayed in Fig. \ref{fig:btbt_hole_particle}. While the hole density illustrates the partial hole transfer from the first to the 5th BTBT monomer from 0~ps to 8~ps, the particle density shows that the electron remains solely on the last monomer for the first 4~ps, transferring a small part of its population to the 7th monomer over 8~ps.
From 12~ps to 20~ps, the majority of the hole population is transferred to the 4th monomer, while the electron is mostly delocalized between the last two BTBT molecules, showing a marginal population of the 6th monomer.
Unfortunately, due to the lack of computational resources, it was not possible to increase the simulation time for the excited-state dynamics trajectories of the BTBT system and prove that our method enables the investigation of the recombination of a CT state to an LE state.
\begin{figure}[b!]
    \centering
    \includegraphics[scale=1.0]{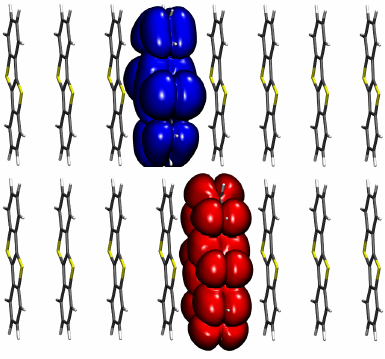}
    \caption{Initial CT state of the simulation of the charge recombination.}
    \label{fig:btbt_recombination}
\end{figure}
Thus, we calculated a single trajectory for the BTBT model system starting from an initial CT state, which was constructed between the 4th and the 5th monomers. The excited-state populations of the BTBT monomers for the simulation of the charge recombination are depicted in Fig. \ref{fig:btbt_recombination}. Within the first  200~fs of the excited-state dynamics, the total population of the CT states decreases by ca. 27\%, while the increased population of the LE states is rapidly delocalized between all the LE states of the BTBT monomers. In the next few ps of the simulation, the CT population is gradually transmitted to the manifold of LE states, resulting in a final excited-state distribution of 39\% LE state and 61\% CT state. 
\begin{figure}[t!]
    \centering
    \includegraphics[width=\linewidth]{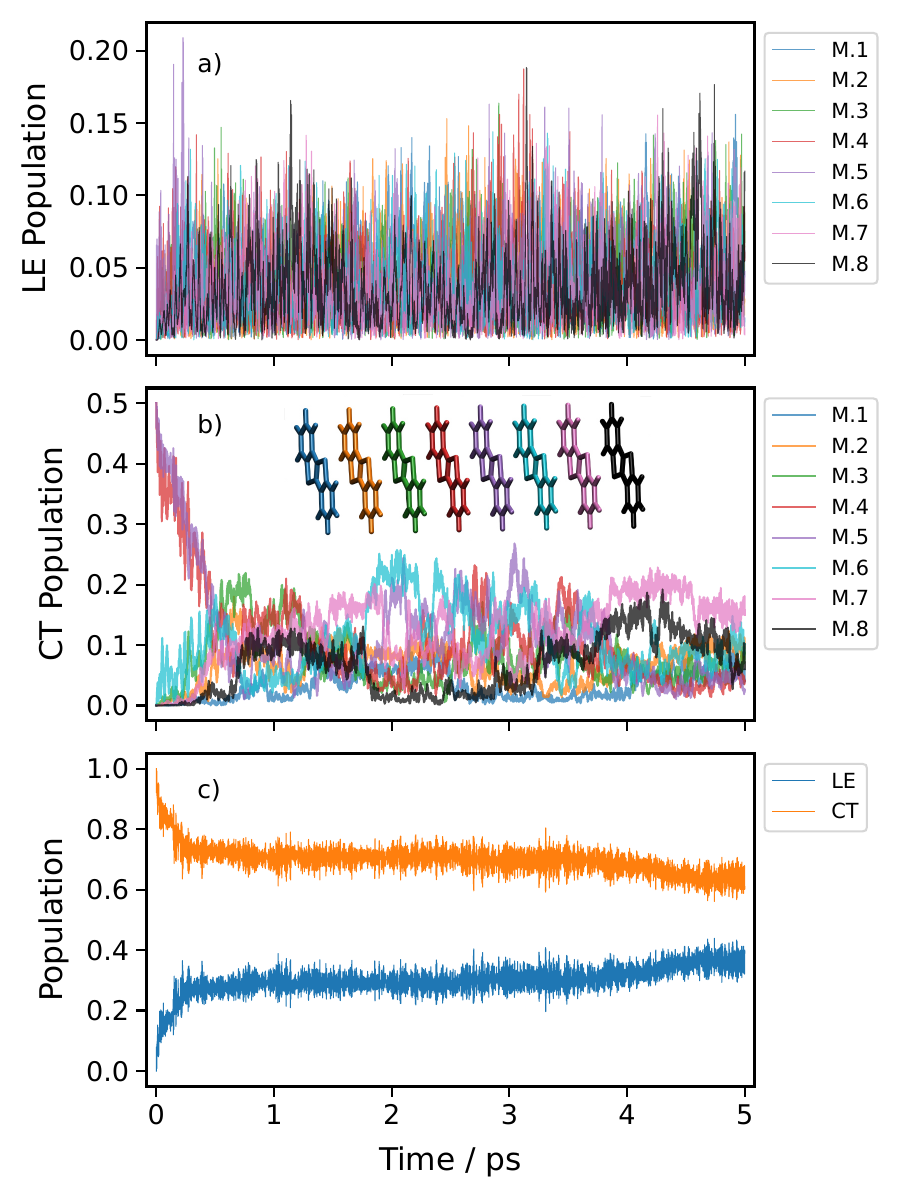}
    \caption{Plot of the a) LE state populations of the monomers, b) CT state populations of the monomers and c) total LE and CT contribution during the simulation.}
    \label{fig:btbt_recombination}
\end{figure}
Another striking feature of the charge recombination simulation is the stark difference of the rate of excitonic transfer between the LE and CT manifolds of the BTBT monomers, which is depicted in Fig. \ref{fig:btbt_recombination} a) and b). This difference results from the disparity in the coupling strength between the respective LE and CT states.

%% file: conclusion.tex
\section{Conclusion and Outlook}\label{sec:conclusion}
We have developed a new methodology to simulate the excited-state dynamics of large molecular aggregates employing the mean-field Ehrenfest approach in the framework of FMO-LC-TDDFTB. To this end, we derived the excited-state gradients and the non-adiabatic couplings for the quasi-diabatic LE and CT states of our FMO-LC-TDDFTB method and implemented the proposed theory in our own software package DIALECT, which is publically available on Github\cite{noauthor_dialect_2024}.

We analyzed the accuracy of the analytical LE and CT gradients by the comparison with the numerical gradients. The results show an average deviation of the excited-state gradients in the range of $10^{-5} E_{\mathrm{h}}/a_0$ (LE) to $10^{-4} E_{\mathrm{h}}/a_0$ (CT), which should be sufficient for the simulation of excited-state dynamics.

The excited-state exciton dynamics of the two-dimensional anthracene system showed the potential applications of our method in simulating the exciton transport in organic semiconductors and nanomaterials. 

The simulation of the charge-transfer dynamics in the model system of 8 BTBT monomers proved the capability of our methodology in investigating the excited-state hole and electron dynamics in organic materials. Our findings indicate an enhanced hole mobility within the BTBT system, consistent with its classification as a p-type semiconductor, thereby showcasing the potential applicability of our methodology. In addition, the excited-state dynamics starting from a neighbouring CT state proved that our approach also facilitates the observation of charge recombination in molecular systems.

Our new methodology opens the possibility to perform fully atomistic real-time simulations of excitonic transfer in large biological systems and organic materials and devices. To this end, we intend to extend the parameterization of the LC-DFTB method to a wider range of elements. Furthermore, we aim to combine the FMO-LC-TDDFTB methodology with the Tavis-Cummings Hamiltonian to calculate the interaction between large molecular systems and microcavities and simulate their excited-state dynamics in the regime of strong light-matter coupling. Additionally, we are working on an ab-initio implementation of our methodology to facilitate the simulation of more complex molecular materials.

%% file: appendix.tex
\begin{appendices}
\section{Determination of Lagrange multipliers}\label{appendix:langrange_multipliers}
A brief summary of the determination of the Lagrange multipliers is presented. For a more detailed derivation, we refer to Ref. \cite{humeniuk_dftbaby_2017}. 

To obtain the Lagrange multipliers $\mathbf{Z}$ and $\mathbf{W}$, the stationary condition of the Lagrange functional in respect to the variation of the molecular orbital coefficients
\begin{equation}
    \sum_{\mu} \frac{\partial L}{\partial C_{\mu p}} C_{\mu q} = 0
\end{equation}
is used. By inserting Eq. (\ref{eq:gradient_lagrangian}) into the above equation, one gets
\begin{align*}\label{eq:appen_lagrange}
    \sum_{\mu} \frac{\partial L}{\partial C_{\mu p}} C_{\mu q} &= \sum_{\mu} \frac{\partial G}{\partial C_{\mu p}} C_{\mu q} + \sum_{ia} Z_{ia} \sum_{\mu} \frac{\partial H_{ia}}{\partial C_{\mu p}} C_{\mu q} \\
    &- \sum_{r \leq s} W_{rs} \sum_{\mu} \frac{\partial S_{rs}}{\partial C_{\mu p}} C_{\mu q} = 0, \numberthis
\end{align*}
where
\begin{equation}
    Q_{pq} = \sum_{\mu} \frac{\partial G}{\partial C_{\mu p}} C_{\mu q}.
\end{equation}
The second and third term terms of Eq. (\ref{eq:appen_lagrange}), which contain $\mathbf{Z}$ and $\mathbf{W}$, are
\begin{align*}\label{eq:appen_z}
\sum_{i a} Z_{i a} \sum_\mu \frac{\partial H_{i a}}{\partial C_{\mu p}} C_{\mu q} & =\sum_{i a} Z_{i a}\left[\left(\delta_{p a} \delta_{q i}+\delta_{p i} \delta q a\right) \epsilon_i \right.  \\
&\left. +\delta_{p \in \mathrm{occ}}(A+B)_{i a, p q}\right] \numberthis \\
&=Z_{q p} \epsilon_q+Z_{p q} \epsilon_p+\delta(p \in \mathrm{occ}) \\
&\times \sum_{i a} Z_{i a}(A+B)_{i a, p q} \numberthis
\end{align*}
and 
\begin{align*}\label{eq:appen_w}
\sum_{r, s, r \leq s} W_{r s} \sum_\mu \frac{\partial S_{r s}}{\partial C_{\mu p}} C_{\mu q} & =\sum_{r, s, r \leq s} W_{r s}\left(\delta_{q s} \delta_{r p}+\delta_{q r} \delta_{s p}\right) \\
& =\left(1+\delta_{p q}\right) W_{p q}. \numberthis
\end{align*}
Inserting the Eqs. (\ref{eq:appen_z}) and (\ref{eq:appen_w}) into (\ref{eq:appen_lagrange}) yields
\begin{align*}\label{eq:appen_zw}
    &Q_{p q}+\left(Z_{q p} \epsilon_q+Z_{p q} \epsilon_p\right)+\delta_{p \in \mathrm{occ}} \sum_{i a} Z_{i a}(A+B)_{i a, p q}\\ &=\left(1+\delta_{p q}\right) W_{p q}, \numberthis
\end{align*}
which is used to determine \textbf{Z}. By subtracting the occupied-virtual 
\begin{align*}
    &Q_{i a}+Z_{i a} \epsilon_i+\sum_{j b}(A+B)_{i a, j b} Z_{j b} \numberthis\\
    &=\left(1+\delta_{i a}\right) W_{i a} \text { for } p \in \text {occ and }  q \in \text{virt}
\end{align*}
from the virtual-occupied
\begin{equation}
    Q_{a i}+Z_{i a} \epsilon_i=\left(1+\delta_{a i}\right) W_{a i} \text { for } p \in \text {virt } \text{and } q \in \text {occ}
\end{equation}
block of Eq. (\ref{eq:appen_zw}), the Z-vector equation is obtained:
\begin{equation}
    \sum_{j b}(A+B)_{i a, j b} Z_{j b}=Q_{a i}-Q_{i a}
\end{equation}
Using the solution of the Z-vector equation, the Lagrange multiplier \textbf{W} is calculated according to
\begin{align}
    W_{ij} &= \frac{1}{1 + \delta_{ij}} \left( Q_{ij} + \sum_{kb} (A+B)_{ij,kb} Z_{kb} \right) \\
    W_{ia} &= W_{ai} = Q_{ai} + Z_{ia} \epsilon_i \\
    W_{ab} &= \frac{1}{1 + \delta_{ab}} Q_{ab},
\end{align}
where the different occupied and virtual parts of $Q_{pq}$ are defined as
\begin{align}
\begin{split}
    Q_{ij} =& \Omega \sum_{c}X_{ic} X_{jc} - 2 \sum_{c} \epsilon_c X_{ic} X_{jc} \\&+ 2 H_{ij}^{+}[T^{\mathrm{v}-\mathrm{v}}] - 2 H_{ij}^{+}[T^{\mathrm{o}-\mathrm{o}}]
\end{split}\\
\begin{split}
    Q_{ia} =& 2\sum_{c}X_{ic} H_{ac}^{+}[\mathbf{X}] +2 H_{ia}^{+}[T^{\mathrm{v}-\mathrm{v}}] \\&- 2H_{ia}^{+}[T^{\mathrm{o}-\mathrm{o}}]  
\end{split}
    \\
    Q_{ai} =& 2\sum_{k} X_{ka} H_{ki}^{+}[\mathbf{X}] \\
    Q_{ab} =& \Omega \sum_{k}X_{ka} X_{kb} + 2 \sum_{k} \epsilon_k X_{kb} X_{ka}, 
\end{align}
with 
\begin{align}
    H_{pq}^{+}[v_{rs}] &= \sum_{r,s} A_{pq,rs} v_{rs} \\
    &=
    \sum_{r,s} \left( 2\left(pq|rs \right) - \left(pr|qs \right) \right) v_{rs}
\end{align}
and 
\begin{align}
    T^{\mathrm{o}-\mathrm{o}} =& \sum_{a} X_{ia} X_{ja} \\
    T^{\mathrm{v}-\mathrm{v}} =& \sum_{i} X_{ia} X_{ib}.
\end{align}

\end{appendices}